\documentclass[10pt,a4paper]{article}
\usepackage[utf8]{inputenc}
\usepackage[english]{babel}

\usepackage{hyperref}

\usepackage{authblk}

\usepackage{feynmp}
\usepackage{amsmath}
\usepackage{amsfonts}
\usepackage{amssymb}

\title{Fab Four Effective Field Theory Treatment}
\author[1,2]{B. Latosh}
\affil[1]{Department of Physics and Astronomy, University of Sussex, Brighton, BN1 9QH, United Kingdom}
\affil[2]{Dubna State University, Universitetskaya str. 19, Dubna 141982, Russia}

\begin{document}

\maketitle

\begin{abstract}
The article addresses the John interaction from Fab Four class of Horndeski models from the effective field theory point of view. Models with this interaction are heavily constrained by gravitational wave speed observations, so it is important to understand, if these constraints hold in the effective field theory framework.
We show that John interaction induces new terms quadratic in curvature at the level of the effective (classical) action. These new terms generate additional low energy scalar and spin-2 gravitational degrees of freedom. Some of them have a non-vanishing decay width and some are ghosts. Discussion of these features is given. 
\end{abstract}

\section{Introduction}

Modified gravity encompasses a broad range of models. Traditionally such models are classified according to the type of modification. For example, models with an additional scalar field are called scalar-tensor gravity; models whose Lagrangian is a continuous function of the scalar curvature $R$ are called $f(R)$ gravity, etc \cite{Berti:2015itd}. One can also classify modified gravity models by their particle content. General Relativity (GR) describes massless spin-2 particles (i.e. gravitons) interacting both with matter and themselves. Modified gravity models change the standard GR content by adding new physical fields \cite{Calmet:2017voc,Hindawi:1995an}. Scalar-tensor models and $f(R)$ gravity serve as the simplest example of such a modification, as they introduce an additional spin-0 particle in the model particle spectrum \cite{DeFelice:2010aj,PhysRevD.39.3159,Wald:1984rg}. In matter of fact, it is possible to map an $f(R)$-gravity model onto a scalar-tensor model with the Brans-Dicke parameter $\omega_\text{BD}=0$ \cite{DeFelice:2010aj,Boisseau:2000pr,Teyssandier:1983zz}. Therefore $f(R)$ gravity and scalar-tensor models should not be treated as completely independent theories. 

Clearly, the simplest way to modify GR is to introduce an additional scalar degree of freedom (DOF). However, 
the new DOF should not be  introduced in an arbitrary manner. The action describing the new DOF must produce second order field equations as a higher derivative action either introduces ghost instabilities or describes additional DOFs (up to a proper reparametrization).

Scalar-tensor models with second order field equations are given by the Horndeski Lagrangian \cite{Horndeski:1974wa} in the Generalized Galileons parameterization \cite{Kobayashi:2011nu}:
\begin{equation}\label{Horndeski_action}
\mathcal{L}=\mathcal{L}_2+\mathcal{L}_3+\mathcal{L}_4+\mathcal{L},
\end{equation}
\begin{align}
\mathcal{L}_2 &=G_2 , \\
\mathcal{L}_3 &=-G_3 \square X , \\
\mathcal{L}_4 &=G_4 R + G_{4X} [(\square \phi)^2 - (\nabla_\mu \nabla_\nu\phi)^2] , \\
\mathcal{L}_5 &=G_5 G_{\mu\nu} \nabla^\mu \nabla^\nu \phi -\cfrac16 G_{5X} \left[(\square \phi)^3 - 3 (\square \phi) (\nabla_\mu \nabla_\nu \phi)^2 + 2(\nabla_\mu \nabla_\nu \phi)^3 \right] .
\end{align}
Here $G_2$, $G_3$, $G_4$, and $G_5$ are functions depending on the scalar field $\phi$ and the standard kinetic term $X=1/2 ~\partial_\mu \phi \partial^\mu \phi$; $G_{4X}$ and $G_{5X}$ are correspondent derivatives with respect to $X$; $G_{\mu\nu}$ is the Einstein tensor. It is worth noting that only $\mathcal{L}_4$ and $\mathcal{L}_5$ describe a non-standard interaction between gravity and the scalar field, while terms $\mathcal{L}_2$ and $\mathcal{L}_3$ describe the scalar field self interaction. 

Horndeski models contain a special subclass called Fab Four \cite{Charmousis:2011bf} which is defined by its ability to screen the cosmological constant. To be exact, Fab Four models completely screen an arbitrary cosmological constant on the Freedman-Robertson-Walker background and such a screening holds even if the cosmological constant experience a finite shift. Fab Four class is given by the following Lagrangian:
\begin{equation}
	\mathcal{L}=\mathcal{L}_\text{John}+\mathcal{L}_\text{George}+\mathcal{L}_\text{Ringo}+\mathcal{L}_\text{Paul} ,
\end{equation}
\begin{align}
\mathcal{L}_\text{John} 	&= V_J(\phi) G_{\mu\nu} \nabla^\mu \phi \nabla^\nu \phi , \\
\mathcal{L}_\text{George} 	&= V_G(\phi) R , \\
\mathcal{L}_\text{Ringo} 	&= V_R(\phi) \hat{G} , \\
\mathcal{L}_\text{Paul} 	&= V_P(\phi) P^{\mu\nu\alpha\beta} \nabla_\mu \phi \nabla_\alpha \phi \nabla_\nu \nabla_\beta \phi .
\end{align}
Here $\hat{G}$ is the Gauss-Bonnet term, $P^{\mu\nu\alpha\beta}=-1/2 ~ \varepsilon^{\alpha\beta\lambda\tau} R_{\lambda\tau\sigma\rho} \varepsilon^{\sigma\rho\mu\nu}$ is the double-dual Riemann tensor, and $V_J$, $V_G$, $V_R$, $V_P$ are interaction potentials. The following features of this class should be noted. The Ringo term alone does not screen the cosmological constant, it just does not ruin the screening. The George term introduces a Brans-Dicke-like coupling which may not be enough to support screening in a particular setting \cite{Charmousis:2011bf}. Therefore only the John and Paul terms drive the screening. Finally, the Paul term demonstrates a pathological behavior in star-like objects \cite{Maselli:2016gxk,Appleby:2015ysa}. Therefore, the John term is the most relevant term for the cosmological constant screening.

A combination of the John term and beyond Fab Four terms allows one to construct a model which may provide an adequate description of both the cosmological expansion while keeping  the cosmological constant small. When the John term is the leading contribution, there is a  screening of the cosmological constant; when the leading contribution is provided by beyond Fab Four terms the model losses its screening properties and develops a small cosmological constant. A particular example of such a model is given in \cite{Starobinsky:2016kua} by the following action (we use different notations):
\begin{equation}\label{Starobinsky_action}
S=\int d^4 x \sqrt{-g} \left[ -\cfrac{1}{16\pi G} (R-2 \Lambda) + \cfrac12 g^{\mu\nu} \nabla_\mu \phi \nabla_\nu \phi +\beta G^{\mu\nu} \nabla_\mu \phi \nabla_\nu\phi \right].
\end{equation}
In full agreement with the aforementioned logic the model describes both inflation and the late-time accelerated expansion of the universe. 

Recent direct detection of gravitational waves (GW) \cite{Abbott:2016blz,Abbott:2016nmj,Abbott:2017vtc,Abbott:2017gyy,Abbott:2017gyy,Abbott:2017oio} and the measurement of the GW speed \cite{TheLIGOScientific:2017qsa,GBM:2017lvd} allow one to establish sever constraints on Horndeski models \cite{Bettoni:2016mij,Ezquiaga:2017ekz}. The authors of \cite{Bettoni:2016mij,Ezquiaga:2017ekz} considered the propagation of tensor perturbations on a cosmological background in Horndeski models and identified them with GWs detected in the terrestrial experiment. Due to the structure of Horndeski models the speed of perturbations strongly depends on $G_4$ and $G_5$, so in order to obtain a model with the constant GW speed one must put the following constraints on the Horndeski parameters:
\begin{align}\label{the_constraints}
	G_{4X}&=0, & G_5=\text{const}.
\end{align}
Constraints \eqref{the_constraints} would rule out the John interaction (and Fab Four in general) from the list of relevant scalar-tensor models.

In this paper we indirectly address constraints \eqref{the_constraints} and their role in the context of the effective field theory treatment of gravity. We claim that although the constraints \eqref{the_constraints} hold for the classical Horndeski action \eqref{Horndeski_action}, one cannot use the action \eqref{Horndeski_action} as a coherent effective action. One must introduce Horndeski interaction at the level of the fundamental action of the model and restore the form of the effective (i.e. classical) gravity action. We present a derivation of such an effective action in the following section. As the effective action does not match the Horndeski action \eqref{Horndeski_action}, the role of the constraints \eqref{the_constraints} should be reconsidered. We discuss this issue in the last section.

\section{Effective Field and John Interaction}\label{main_section}

The standard Effective Field Theory (EFT) technique is based on the following premise \cite{Burgess:2003jk}. Let us assume that one has a physical system containing heavy $\mathit{h}$ and light $\mathit{l}$ degrees of freedom described by a fundamental action $\mathcal{A}[\mathit{l},\mathit{h}]$. In the low energy regime, i.e. when energies are below the heavy degree of freedom (DOF) mass scale, the system is described by an effective action $\Gamma[\mathit{l}]$ given in terms of the light DOF only. In order to obtain the effective action one must integrate out the heavy DOF:
\begin{equation}
\int \mathcal{D}[\mathit{l}] \exp\left[ i~ \Gamma[\mathit{l}] ~ \right]=\int \mathcal{D}[\mathit{l}] \mathcal{D}[\mathit{h}] \exp\left[i~\mathcal{A}[\mathit{l},\mathit{h}] ~\right] ~.
\end{equation}
If the fundamental action is unknown, one can restore its form, as it must contain all terms permitted by general covariance, conservation laws, and other fundamental physical principles. 

A similar logic holds for gravity models \cite{Donoghue:1994dn,Donoghue:2012zc}. One assumes, that gravity is described by some fundamental action $\mathcal{A}[\mathfrak{g}]$ which is given in terms of the metric $\mathfrak{g}$ describing behavior of the true quantum gravitons propagating over some background spacetime $g$. In order to obtain an effective action, one must integrate out all quantum gravitons:
\begin{equation}
\exp\left[ i \Gamma[g_{\mu\nu}] \right] = \int \mathcal{D}[\mathfrak{g}] \exp\left[ i \mathcal{A}[g_{\mu\nu}+\mathfrak{g}_{\mu\nu}] \right].
\end{equation}
The effective action $\Gamma[g]$ is given in terms of the classical metric $g_{\mu\nu}$ which is generated by the underlying dynamic of quantum gravitons. Let us emphasis, that in such an implication one must integrate out light DOF unlike the before-mentioned case. Moreover, the fundamental action $\mathcal{A}$ must contain data on the matter content of the Universe, as it describes both gravity and its interaction with matter.

Despite the fact that the fundamental action for gravity is unknown, one can restore the form of the effective action. One must include $R^2$ and $R_{\mu\nu}^2$ terms in $\Gamma$, as the correspondent operators are generated at the level of the first matter loop \cite{tHooft:1974toh}. Moreover, one must include nonlocal operators \cite{Donoghue:2014yha,Calmet:2015dpa,Alexeyev:2017scq}, as one can sum up an infinite series of matter loops in the graviton propagator. We do not discuss this feature in details, as it lies beyond the scope of this paper and it was covered in details in \cite{Donoghue:1994dn,Donoghue:2012zc,Donoghue:2014yha,Calmet:2017voc,Alexeyev:2017scq}.
The standard approach to modified gravity is to consider only the classical action describing gravity without respect to the underlying quantum dynamic of the gravitational field  \cite{Berti:2015itd,Bettoni:2016mij,Ezquiaga:2017ekz}. This approach is coherent and appears to be fruitful in modified gravity studies. Within the EFT framework one is obliged to consider the classical action as the effective action, so it cannot be taken arbitrary. 

In such a way we claim that if the effective (classical) action of gravity contains Horndeski interactions, then the fundamental action for gravity also must contain a Horndeski sector and vice versa. In this paper we address a particular action \eqref{Starobinsky_action} as it is well motivated. We consider \eqref{Starobinsky_action} as a part of the fundamental gravity action. In order to derive interaction rules we expand the action over a flat background:
\begin{align}\label{Starobinsky_linear_action}
&\mathcal{A}=\int d^4 x \left[ -\cfrac12 h^{\mu\nu} \mathcal{O}_{\mu\nu\alpha\beta} h^{\alpha\beta} - \cfrac12 \phi \square \phi -\cfrac{\kappa}{4} h^{\mu\nu} C_{\mu\nu\alpha\beta} \partial^\alpha \phi \partial^\beta \phi \right .  \\
&\left. +\cfrac{\kappa}{2} \beta \left[ \partial_\mu \partial^\sigma h_{\nu\sigma} +\partial_\nu \partial^\sigma h_{\mu\sigma} - \partial_\mu \partial_\nu h - \square h_{\mu\nu} - \eta_{\mu\nu} (\partial_\alpha\partial_\beta h^{\alpha\beta} - \square h) \right] \partial^\mu \phi \partial^\nu\phi \right] . \nonumber
\end{align}
Correspondent Feynman rules are presented in the Appendix. An expansion about the flat background is due, as we are interested in the local effects. The presence of the cosmological constant can be neglected for the sake of simplicity, as it strongly affects only large-scale physics.

Following the standard procedure presented in \cite{Donoghue:1994dn,Donoghue:2012zc,Burgess:2003jk} one can calculate the effective action based on the form of the fundamental action \eqref{Starobinsky_action}. The effective action is given by the following:
\begin{align}\label{the_effective_action}
& \Gamma=\int d^4 x \sqrt{-g} \left[ -\cfrac{1}{16\pi G} R + \cfrac12 g^{\mu\nu} \nabla_\mu \phi \nabla_\nu \phi + \beta G^{\mu\nu} \nabla_\mu \phi \nabla_\nu\phi \right. \\
 &\left. + c_1 R^2 + c_2 R_{\mu\nu}^2 + \bar{c}_1 R \beta\square R + \bar{c}_2 R_{\mu\nu} \beta\square R^{\mu\nu} + \bar{\bar{c}}_1 R (\beta \square)^2 R + \bar{\bar{c}}_2 R_{\mu\nu} (\beta \square)^2 R^{\mu\nu} \right] \nonumber .
\end{align}
Here $c_1$, $c_2$, $\bar{c}_1$, $\bar{c}_2$, $\bar{\bar{c}}_1$, and $\bar{\bar{c}}_2$ are unknown dimensionless constants. First three terms in \eqref{the_effective_action} appear in the effective action due to the fact that they present in the fundamental action. Terms $R^2$ and $R_{\mu\nu}^2$ appear due to the first matter loop associated with the standard interaction between gravity and matter \cite{tHooft:1974toh}. Terms with $\beta \square$ appear due to the presence of the John interaction and its contribution to the first order matter loop (correspondent diagrams are given in the Appendix \eqref{loop_two},\eqref{loop_three}). Strictly speaking, the part of action \eqref{the_effective_action} containing the scalar field belongs to the Horndeski class, but the action also has terms missing in the Horndeski action. This implies that existing constraints \eqref{the_constraints} cannot be directly applied in such a framework. The dynamic of tensor perturbations over a cosmological background described by \eqref{the_effective_action} differs from the one described by the Horndeski action, so the proper constraints should be found. This however doesn't ruin the constraint's applicability outside the EFT framework and they must be used to constraint Horndeski models in the original modified gravity framework. The effective action \eqref{the_effective_action} itself requires a further analysis.

In full analogy with the classical results \cite{Stelle:1976gc,Stelle:1977ry} (see also \cite{Accioly:2000nm} for a more detailed derivation) higher derivative terms change the content of the model. Terms $R^2$ and $R_{\mu\nu}^2$ introduce additional massive spin-2 and spin-0 degrees of freedom. Following the algorithm presented in \cite{Accioly:2000nm} one can calculate a propagator of gravity modes given by the effective action \eqref{the_effective_action}:
\begin{align}\label{the_propagator}
G_{\mu\nu\alpha\beta} (k) &= \cfrac{1}{k^2} \left[ P^1_{\mu\nu\alpha\beta} + \cfrac{P^2_{\mu\nu\alpha\beta}}{1+\cfrac12 ~\kappa^2 k^2 (c_2-\bar{c}_2 \beta k^2 + \bar{\bar{c}}_2 (\beta k^2)^2)} \right.
\end{align}
\begin{align}
 &-\cfrac12 \cfrac{P^0_{\mu\nu\alpha\beta} +\bar{\bar{P}}^0_{\mu\nu\alpha\beta} }{1- \kappa^2 k^2 ( 3 c_1 +c_2 - (3\bar{c}_1 + \bar{c}_2 )\beta k^2 + (3 \bar{\bar{c}}_1+\bar{\bar{c}}_2) (\beta k^2)^2) } \nonumber \\
& \left. -\cfrac12 \cfrac{\left[3-4 (1- \kappa^2 k^2 ( 3 c_1 +c_2) - (3\bar{c}_1 + \bar{c}_2 )\beta k^2 + (3 \bar{\bar{c}}_1+\bar{\bar{c}}_2) (\beta k^2)^2) \right] }{1- \kappa^2 k^2 ( 3 c_1 +c_2 - (3\bar{c}_1 + \bar{c}_2 )\beta k^2 + (3 \bar{\bar{c}}_1+\bar{\bar{c}}_2) (\beta k^2)^2) } \bar{P}^0_{\mu\nu\alpha\beta}  \right] ~. \nonumber
\end{align}
In this expression operators $P^2$, $P^0$, $\bar{P}^0$, and $\bar{\bar{P}}^0$ are taken directly from \cite{Accioly:2000nm}. Each pole in the propagator corresponds to a new particle state. Because of the similarity between action \eqref{the_effective_action} and the well-known Stelle action \cite{Stelle:1976gc,Stelle:1977ry,Accioly:2000nm}, it can be seen that propagator \eqref{the_propagator} describes additional scalar and spin-2 particles. The denominator of the propagator is a forth order polynomial in $k^2$, which means that the propagator has four complex poles.
With the use of the fundamental theorem of algebra one can establish, that poles for the spin-0 mode are located in points $k^2 =\pm m_0^2$, $\pm i m_0^2$, where $m_0$ is a real constant; a similar statement holds for the spin-2 mode: $k^2 = \pm m_2^2$, $\pm i m_2^2$, where $m_2$ is another real constant. These poles corresponds to massive scalar, massive spin-2 particle, massive scalar ghost, massive spin-2 ghost, two massive spin-0 particles with non-zero decay width, and two massive spin-0 particles with non-zero decay width.

These results follow the standard  EFT logic \cite{Donoghue:1994dn,Donoghue:2012zc,Donoghue:2014yha,Calmet:2017voc}. The presence of ghost states in such models is a well-known feature. The appearance of new massive states is also a typical feature of EFT models discovered in the classical papers \cite{Stelle:1976gc,Stelle:1977ry}. Therefore the effective model has the standard EFT features and can be considered alongside the regular EFT models.

\section{Discussion and conclusion}\label{discussion_and_conclusion}

In this paper we have used effective field theory techniques to restore a form of the classical gravity action. We used a particular Horndeski model \cite{Starobinsky:2016kua} as a part of the fundamental gravity action in order to generate the effective action. Such an approach is necessary to study scalar-tensor gravity models and modified gravity in general. We obtained the effective action \eqref{the_effective_action} generated by the fundamental action \eqref{Starobinsky_action}. This fundamental action contains higher derivative terms, which leads to the following consequences. 

First of all, the constraints \eqref{the_constraints} obtained in \cite{Bettoni:2016mij,Ezquiaga:2017ekz} cannot be used within EFT framework. These constraints \eqref{the_constraints} were obtained from a study of tensor perturbations in Horndeski models, however the effective action \eqref{the_effective_action} differs from the Horndeski action \eqref{Horndeski_action} and thus the dynamics of tensor perturbations is also different. Therefore the constraints \eqref{the_constraints} do not hold in the EFT framework, although this does not affect their relevance for the classical modified gravity framework.

Secondly, we analyzed the content of the effective action. The particle spectrum of the model is given by the propagator of the low energy gravity perturbations \eqref{the_propagator}. The new low energy degrees of freedom are a massive scalar particle, a massive spin-2 particle, a massive scalar ghost, a massive spin-2 ghost, two massive scalar particles with non-vanishing decay width, and two massive spin-2 particles with non-vanishing decay width. The presence of ghost states and states with non-zero decay width is typical for models of such kind \cite{Stelle:1977ry,Donoghue:1994dn,Donoghue:2012zc,Calmet:2015dpa,Alexeyev:2017scq,Calmet:2017qqa}. We prefer not to draw any conclusion on the relevance of the model based on the fact that it contains ghosts, as the issue is typical for a number of before-mentioned effective field models of gravity. We, however, argue that the presence of new gravitational degrees of freedom must affect late stages of GW production during the last stages of binary systems coalescence, as was shown in \cite{Calmet:2018qwg}.

Summarizing all the results we make the following conclusions. The existence of non-trivial Horndeski interaction at the level of the fundamental action induces non-trivial corrections to low energy gravitational phenomena. The effective model discussed in this paper provides the simplest example of such phenomena. This model shares problems typical of all gravitational EFT. Finally, some well-known constraints on Horndeski models \eqref{the_constraints} cannot be applied to it. We finish by emphasizing that this model seems to be a rather special modification of the standard gravity EFT. 

\section*{Acknowledgements}
This work was supported by Russian Foundation for Basic Research via grant RFBR 16-02-00682.

\section{Appendix}

Operator $\mathcal{O}$ used in \eqref{Starobinsky_linear_action} is given by the following expression:
\begin{align}
& \mathcal{O}_{\mu\nu\alpha\beta} = \cfrac12 \left(\eta_{\mu\alpha}\eta_{\nu\beta} + \eta_{\mu\beta} \eta_{\nu\alpha}  \right) \square - \eta_{\mu\nu}\eta_{\alpha\beta} \square +(\partial_\mu \partial_\nu \eta_{\alpha\beta} + \partial_\alpha\partial_\beta \eta_{\mu\nu}) \nonumber \\
& -\cfrac12 \left( \partial_\alpha\partial_\mu \eta_{\beta\nu} + \partial_\alpha\partial_\nu \eta_{\beta\mu}+ \partial_\beta\partial_\mu \eta_{\alpha\nu}+\partial_\beta\partial_\nu \eta_{\alpha\mu} \right) .
\end{align}

Action \eqref{Starobinsky_linear_action} generates the following Feynman rules for propagators:
\begin{align}
  \begin{gathered}
    \begin{fmffile}{graviton_propagator}
      \begin{fmfgraph}(60,60) 
        \fmfleft{i}
        \fmfright{o}
        \fmf{dbl_wiggly}{i,o}
      \end{fmfgraph}
    \end{fmffile}
  \end{gathered}
  &= \cfrac{i}{2} \cfrac{C_{\mu\nu\alpha\beta}}{k^2} , &
  \begin{gathered}
    \begin{fmffile}{scalar_propagator}
      \begin{fmfgraph}(60,60)
        \fmfleft{i}
        \fmfright{o}
        \fmf{dashes}{i,o}
      \end{fmfgraph}
    \end{fmffile}
  \end{gathered}
  &= ~\cfrac{i}{k^2} .
\end{align}
Here $C_{\mu\nu\alpha\beta}$ is defined as follows:
\begin{equation}
C_{\mu\nu\alpha\beta} = \eta_{\mu\alpha} \eta_{\nu\beta} + \eta_{\mu\beta} \eta_{\nu\alpha} - \eta_{\mu\nu} \eta_{\alpha\beta}.
\end{equation}

For the standard interaction between gravity and the scalar field the correspondent rule reads:
\begin{align}
  \begin{gathered}
    \begin{fmffile}{scalar_energy-momenum_tensor_kinetic_contribution}
      \begin{fmfgraph*}(60,60)
        \fmfleft{i1,i2}
        \fmfright{o}
        \fmf{dashes_arrow,label=$p$}{i1,v}
        \fmf{dashes_arrow,label=$q$}{i2,v}
        \fmf{dbl_wiggly,tension=2}{o,v}
        \fmfdot{v}
        \fmflabel{$\mu\nu$}{o}
      \end{fmfgraph*}
    \end{fmffile}
  \end{gathered} \hspace{.7cm}
  &= i\cfrac{\kappa}{4}  C_{\mu\nu\alpha\beta} p^\alpha q^\beta .
\end{align}

For John interaction the correspondent rule reads:
\begin{align}
  \begin{gathered}
    \begin{fmffile}{John_interaction}
      \begin{fmfgraph*}(60,60)
        \fmfleft{i1,i2}
        \fmfright{o}
        \fmf{dashes_arrow,label=$p$}{i1,v}
        \fmf{dashes_arrow,label=$q$,label.side=left}{i2,v}
        \fmf{dbl_wiggly,tension=2,label=$k$}{v,o}
        \fmfv{decor.shape=square,decor.filled=full,decor.size=2thick}{v}
        \fmflabel{$\mu\nu$}{o}
      \end{fmfgraph*}
    \end{fmffile}
  \end{gathered} \hspace{.7cm}
  &= i ~ \cfrac{\kappa}{2} ~ \beta ~ k^2 ~ \mathcal{M}_{\mu\nu\alpha\beta}(k) p^\alpha q^\beta ,
\end{align}  
\begin{align}
  \mathcal{M}_{\mu\nu\alpha\beta} (k) &\overset{\text{def}}{=}\eta_{\mu\nu} \eta_{\alpha\beta} - I_{\mu\nu\alpha\beta}  - (\omega_{\mu\nu} \eta_{\alpha\beta} + \eta_{\mu\nu} \omega_{\alpha\beta}) \nonumber \\
  &+\cfrac12 \left( \omega_{\mu\alpha} \eta_{\nu\beta} + \omega_{\mu\beta} \eta_{\nu\alpha} + \omega_{\nu\alpha} \eta_{\mu\beta} + \omega_{\nu\beta} \eta_{\mu\alpha} \right), \\
  \omega_{\mu\nu}(k) &\overset{\text{def}}{=} \cfrac{k_\mu k_\nu}{k^2} .
\end{align}

The existence of the John interaction vertex provides two new one-loop level diagrams. In such a way there are three one loop level diagrams generated by action \eqref{Starobinsky_action}. Their divergent parts are evaluated in the dimensional-regularization scheme and read:
\begin{align}
  \begin{gathered}
    \begin{fmffile}{graviton_scalar-field_polarization_operator}
      \begin{fmfgraph*}(80,80)
        \fmfleft{i}
        \fmfright{o}
        \fmf{dbl_wiggly,tension=1.5}{i,vl}
        \fmf{dbl_wiggly,tension=1.5}{vr,o}
        \fmf{phantom}{vl,vr}
        \fmfdot{vl,vr}
        \fmffreeze
        \fmf{dashes,left}{vl,vr}
        \fmf{dashes,left}{vr,vl}
      \end{fmfgraph*}
    \end{fmffile}
  \end{gathered}
  &\to \cfrac{i}{1920\pi^2} \left( R_{\mu\nu} R^{\mu\nu} +\cfrac{11}{4} R^2 \right), \label{loop_one} \\
  \begin{gathered}
    \begin{fmffile}{graviton_John-scalar_polarization_operator}
      \begin{fmfgraph*}(80,80)
        \fmfleft{i}
        \fmfright{o}
		\fmf{dbl_wiggly,tension=1.5}{i,vl}
        \fmf{dbl_wiggly,tension=1.5}{vr,o}
        \fmf{phantom}{vl,vr}
        \fmfdot{vl}
        \fmfv{decor.shape=square,decor.filled=full,decor.size=2thick}{vr}
        \fmffreeze
        \fmf{dashes,left}{vl,vr}
        \fmf{dashes,left}{vr,vl}
      \end{fmfgraph*}
    \end{fmffile}
  \end{gathered}
  &\to\cfrac{i}{1920\pi^2} \left( R_{\mu\nu} (\beta \square) R^{\mu\nu} +\cfrac{11}{4} R (\beta \square) R \right) , \label{loop_two} \\
\begin{gathered}
    \begin{fmffile}{graviton_John_polarization_operator}
      \begin{fmfgraph*}(80,80)
        \fmfleft{i}
        \fmfright{o}
		\fmf{dbl_wiggly,tension=1.5}{i,vl}
        \fmf{dbl_wiggly,tension=1.5}{vr,o}
        \fmf{phantom}{vl,vr}
        \fmfv{decor.shape=square,decor.filled=full,decor.size=2thick}{vl}
        \fmfv{decor.shape=square,decor.filled=full,decor.size=2thick}{vr}
        \fmffreeze
        \fmf{dashes,left}{vl,vr}
        \fmf{dashes,left}{vr,vl}
      \end{fmfgraph*}
    \end{fmffile}
  \end{gathered}
  &\to\cfrac{i}{1920\pi^2} \left( R_{\mu\nu} (\beta \square)^2 R^{\mu\nu} +\cfrac{11}{4} R (\beta \square)^2 R \right). \label{loop_three}
\end{align}

\bibliographystyle{unsrturl}
\bibliography{Biblio.bib}

\end{document}